\documentclass[amsfonts,amsmath,prd,twocolumn,nofootinbib,a4paper,eqsecnum,superscriptaddress]{revtex4-2}
\newcommand{\beq}{\begin{equation}}
\newcommand{\eeq}{\end{equation}}

\newcommand{\bsp}{\begin{split}}

\usepackage{epsfig,bbm,cancel}
\usepackage[normalem]{ulem}
\usepackage[colorlinks=true]{hyperref}
\usepackage{latexsym}
\usepackage[utf8]{inputenc}
\usepackage{amsmath}
\usepackage{amsfonts}
\usepackage{amssymb}
\usepackage{booktabs}
\usepackage{array}
\usepackage{tabularx}
\usepackage{multirow}
\usepackage{longtable}
\usepackage{ragged2e}
\usepackage{braket,booktabs}
\usepackage{braket,amsmath}
\usepackage{xcolor}
\usepackage{graphicx}
\usepackage{orcidlink} 

\begin{document}

\title{New Universal Relations for Magnetized Neutron Stars}

\author{Byon N. Jayawiguna \orcidlink{0000-0001-5920-8701}}
\email{nugrahabyon312@gmail.com}
\affiliation{High Energy Physics Theory Group, Department of Physics, Faculty of Science, Chulalongkorn University, Bangkok 10330, Thailand}
\affiliation{Department of Physics, University of Virginia, Charlottesville, Virginia 22904, USA}

\author{Siddarth Ajith \orcidlink{0000-0002-3360-5810}}
\email{sa4fb@virginia.edu}
\affiliation{Department of Physics, University of Virginia, Charlottesville, Virginia 22904, USA}

\author{Takuya Katagiri  \orcidlink{0000-0002-3755-3093}}
\email{takuya.katagiri@uniroma1.it}
\affiliation{Dipartimento di Fisica, Sapienza Università di Roma, Piazzale Aldo Moro 5, 00185, Roma, Italy}
\affiliation{INFN, Sezione di Roma, Piazzale Aldo Moro 2, 00185, Roma, Italy}

\author{Kent Yagi \orcidlink{0000-0002-0642-5363}}
\email{ky5t@virginia.edu}
\affiliation{Department of Physics, University of Virginia, Charlottesville, Virginia 22904, USA}

\author{Piyabut Burikham \orcidlink{0000-0003-3059-8747}}
\email{piyabut@gmail.com}
\affiliation{High Energy Physics Theory Group, Department of Physics, Faculty of Science, Chulalongkorn University, Bangkok 10330, Thailand}

\def\changenote#1{\footnote{\bf #1}}

\begin{abstract}

Unlike the neutron-star mass--radius relation, which depends sensitively on the internal stellar structure through the equation of state of dense nuclear matter, certain neutron-star properties obey approximately universal relations that exhibit only weak dependence on the equation of state. In this paper, we explore new universal relations for magnetized neutron stars. We focus on a purely poloidal dipolar magnetic-field configuration and treat the effects of the magnetic field perturbatively to second order in the field strength. After appropriately normalizing the relevant quantities by the stellar mass and the magnetic-field strength at the pole, we first identify an exact universal relation between the magnetic dipole moment and stellar compactness. We then find approximate universal relations among the magnetic dipole moment, magnetically-induced stellar quadrupole moment, and ellipticity, with fractional variations due to the equation of state at the level of $\mathcal{O}(10\%)$. We support these numerical findings with analytic estimates for Newtonian polytropes, which reproduce the observed behavior. Among the newly discovered relations, the one connecting the magnetic dipole moment and stellar quadrupole moment may be particularly useful for the analysis of gravitational-wave signals from binaries containing magnetized neutron stars.
\end{abstract}

\maketitle
\thispagestyle{empty}
\setcounter{page}{1}

\section{Introduction}

Neutron stars (NSs) probe the interface between nuclear physics and astrophysics, providing a natural laboratory for studying matter at supranuclear densities under strong gravitational fields \cite{Lattimer:2000nx,Burgay:2003jj,Steiner:2010fz,Ozel:2016oaf,Rezzolla:2016nxn,Annala:2017llu,Most:2018hfd}. Their macroscopic properties are determined by their internal structure, which is governed by the equation of state (EoS) relating pressure and energy density. Radio observations of massive pulsars have ruled out EoSs that cannot support a $2M_\odot$ NS~\cite{1.97NS,2.01NS,NANOGrav:2019jur,Saffer:2024tlb}. Measurements of stellar radii with X-rays~\cite{Riley:2019yda,Miller:2019cac,Miller:2021qha,Raaijmakers:2019qny,Raaijmakers:2021uju,Yunes:2022ldq} and tidal deformabilities with gravitational waves~\cite{LIGOScientific:2017vwq,LIGOScientific:2018cki,Yunes:2022ldq} have further constrained viable EoSs. Other observations, such as quasi-periodic oscillations in gamma-ray bursts~\cite{Guedes:2024zkh}, may also provide complementary constraints on the EoS.

Unlike quantities such as the mass--radius relation or tidal deformability, which depend strongly on the underlying EoS, there exist (quasi-)universal relations that are largely insensitive to the EoS~\cite{Yagi:2016bkt}. A prominent example is the I-Love-Q relation~\cite{Yagi:2013bca,Yagi:2013awa} among the moment of inertia, tidal deformability, and quadrupole moment. Such universal relations have found numerous applications in astrophysics~\cite{Yagi:2013bca,Yagi:2013awa,Newton:2016weo,Xie:2022brn}, nuclear physics~\cite{Yagi:2015pkc,Yagi:2016qmr,LIGOScientific:2018cki}, gravitational physics~\cite{Yagi:2013bca,Yagi:2013awa,Gupta:2017vsl,Silva:2020acr,Saffer:2021gak,Pan:2022gwf,Jayawiguna:2025edl}, and cosmology~\cite{Chatterjee:2021xrm}. For example, the relation between tidal deformability and quadrupole moment allows the latter to be expressed in terms of the former in gravitational-wave models of binary NS inspirals, thereby reducing the dimensionality of the parameter space. This partially breaks parameter degeneracies and improves the measurability of other quantities, such as stellar spins~\cite{Yagi:2016bkt,Yagi:2013bca,Yagi:2013awa}.

Most studies of universal relations have focused on non-magnetized NSs. However, some NSs, known as magnetars, can possess extremely strong magnetic fields~\cite{kulkarni1993,murakami}. References~\cite{kouv1,kouv2,duncan,thompson1993,thompson1995,thompson1996} suggest that several observed pulsars are ultramagnetized, with magnetic-field strengths reaching $10^{14}$--$10^{15}$ G \cite{mereghetti,taylor}. Magnetars have been proposed as central engines for a variety of energetic astrophysical phenomena, including short gamma-ray bursts~\cite{Gao:2022iwn,Rueda:2025lgq} and fast radio bursts~\cite{CHIMEFRB:2020abu,Bochenek:2020zxn,Mereghetti:2020unm,Tavani:2020adq,Insight-HXMTTeam:2020dmu,Ridnaia:2020gcv}. They may also be formed following the merger of two weakly magnetized NSs~\cite{Kiuchi:2026pgb}.

Several formalisms have been developed to construct equilibrium configurations of magnetized stars and study their dynamics, including precession, oscillations, and gravitational-wave emission \cite{Ioka:2003dd,Bonazzola:1995rb,Cutler:2002nw,Ioka:2000yb}. Konno \textit{et al}.~\cite{Konno:1999zv,Konno:2000tw} developed a perturbative framework for constructing static equilibrium configurations of magnetized NSs by extending the slow-rotation formalism of Hartle and Thorne~\cite{Hartle:1967he,Hartle:1968si}. Their approach may be viewed as the relativistic counterpart of the Newtonian framework developed in~\cite{chan,ferraro}. The formalism was subsequently generalized to slowly rotating magnetized NSs~\cite{Konno:2001jr}. A fully relativistic numerical treatment of rotating magnetized NSs was presented in~\cite{Bocquet:1995je}. Equilibrium configurations with purely poloidal, twisted-torus, and other magnetic-field geometries have also been studied extensively; see, e.g., \cite{Asai:2015qha,Ciolfi:2010td,deSouza:2017qwv,Ciolfi:2009bv}.

Only a few studies have investigated universal relations in magnetized NSs. Haskell \textit{et al}.~\cite{Haskell:2013vha} examined how magnetic fields affect the universality of the relation between the moment of inertia and quadrupole moment. In this case, the total quadrupole moment receives contributions from both spin-induced and magnetically induced deformations of spacetime. Following the original I-Love-Q studies~\cite{Yagi:2013bca,Yagi:2013awa}, the authors normalized the quadrupole moment using the stellar mass and spin angular momentum. Considering purely poloidal, purely toroidal, and twisted-torus magnetic-field configurations, they found that the universality generally improves as the stellar rotation rate increases. Physically, this occurs because magnetic effects become less important relative to rotational effects at higher spin frequencies. Conversely, the universality deteriorates in the slow-rotation regime, where magnetic deformations are more pronounced. This work was later extended in~\cite{Zhu:2020imp}, which focused on magnetic corrections to the tidal deformability for a purely poloidal dipolar magnetic-field configuration.

In this paper, we revisit the problem and search for new universal relations in magnetized NSs. We focus on a purely poloidal dipolar magnetic-field configuration as in~\cite{Konno:1999zv,Zhu:2020imp}, and treat magnetic effects perturbatively, retaining terms up to second order in the magnetic-field strength following~\cite{Konno:1999zv}. In particular, we investigate relations among the stellar compactness, magnetic dipole moment, 
magnetically-induced quadrupole moment, and ellipticity. We normalize the latter three quantities using the stellar mass and the magnetic-field strength at the stellar pole. First, we show that an exact universal relation exists between the normalized magnetic dipole moment and compactness. We then demonstrate the existence of approximate universal relations among the magnetic dipole moment, quadrupole moment, and ellipticity, with EoS variations at the $\mathcal{O}(10\%)$ level. To support our numerical results, we derive analytic estimates in the Newtonian limit for polytropic EoSs. The universal relations discovered here, particularly the one connecting the magnetic dipole moment and stellar quadrupole moment, may prove useful for gravitational-wave analyses of binaries containing magnetized NSs, since these quantities enter the waveform phase at the same leading post-Newtonian order~\cite{Ioka:2000yb,poisson-quadrupole}\footnote{\label{footnote} Poisson~\cite{poisson-quadrupole} analyzes the impact of the spin-induced quadrupole moment on gravitational-wave emission under the assumption that the spinning body is axisymmetric about the rotation axis $\hat{s}$. This symmetry leads to the quadrupole structure given in Eq.~(6) of~\cite{poisson-quadrupole}. Although derived for spin-induced deformation, the formalism for the non-precessing case is general and it also applies to magnetically induced quadrupole moments when $\hat{s}$ is reinterpreted as the direction of the magnetic field.
In our case, we extract the magnetically induced quadrupole moment using the Hartle-Thorne framework~\cite{Hartle:1967he,Hartle:1968si} as implemented for magnetic deformations in~\cite{Konno:1999zv}. Consequently, the structural form of the magnetically induced quadrupole moment mirrors that of the spin-induced one. More broadly, the leading contribution of any quadrupole moment (that is independent of the binary separation) to the gravitational-wave phase enters at second post-Newtonian order, as shown for generic axisymmetric sources with arbitrary multipole moments in~\cite{Ryan:1995wh,Ryan:1997hg}. }.

The remainder of this paper is organized as follows. In Sec.~\ref{sec:MagnetizedNSs}, we present the perturbative framework used to construct magnetized NSs. Section~\ref{sec:Results} contains our main results on the approximate universality of magnetized NSs, together with analytic calculations in the Newtonian limit. We summarize our findings and discuss future directions in Sec.~\ref{sec:summary}. Appendix~\ref{app:pert_eqs} presents equations to be solved in the perturbative framework, together with the boundary conditions imposed. Appendix~\ref{app:numerical} describes the numerical procedures used to solve the interior problem. Throughout this paper, we adopt geometric units with $c=G=1$.

\section{Magnetized Neutron Stars}	
\label{sec:MagnetizedNSs}

We construct magnetized NSs within the perturbative framework about the magnetic field strength.

\subsection{Perturbative construction}
Following the manner in~\cite{Konno:1999zv}, we derive perturbation equations from the Einstein and Maxwell equations order by order in $\epsilon$, a bookkeeping parameter to count the power in the magnetic field strength. 


\subsubsection{$\mathcal{O}(\epsilon^0)$: Background Spherically-symmetric Configurations}

The background geometry at $\mathcal{O}(\epsilon^0)$ is assumed to be static and spherically symmetric. The metric,~$g_{\mu\nu}$, is given by 
\begin{equation}
g_{\mu\nu}dx^\mu dx^\nu=-e^{\nu(r)}dt^2 + e^{\psi(r)}dr^2+r^2 d\theta^2+r^2\sin^2\theta d\phi^2.
\end{equation}

We model matter as a perfect fluid with the stress-energy

\begin{equation}
T_{\mu\nu}=(\rho+p) u_{\mu}u_{\nu} + p g_{\mu\nu},
\end{equation} 
where $ p $, $ \rho $, and $ u_{\mu} $, are the pressure, total energy density, and unit timelike four-velocity,  respectively.

Let us first provide the TOV equations at $\mathcal{O}(\epsilon^0)$.
The ($t,t $) and $ (r,r) $ components of the Einstein 
equations yield
\begin{eqnarray}
\label{masseq}
\frac{dm}{dr}&=& 4\pi r^2 \rho, \\ 
\label{eq:nu}
\frac{d\nu}{dr} &=& \frac{8\pi r^3 p + 2m}{r(r-2m)}.    
\end{eqnarray}
Meanwhile, the radial component of the conservation of the background energy-momentum tensor, $ \nabla_{\mu}T^{\mu r}=0$, yields
\begin{equation}
\label{tov}
\frac{dp}{dr}=-(\rho+p) \frac{m+4\pi p r^3}{r(r-2m)}.
\end{equation}

We next describe the boundary conditions. The interior solutions to Eqs.~\eqref{masseq}--\eqref{tov} are matched to the exterior solutions at the stellar surface $R$. The stellar radius $R$ is defined by $p(R)=0$ while the stellar mass is given by $M=m(R)$. The exterior spacetime is described by the Schwarzschild solution:
\begin{equation}
e^{\nu(r)}=e^{-\psi(r)}=1-\frac{2M}{r},~~~\textrm{for $ r>R. $}
\end{equation}
The boundary conditions at the surface are given by
\begin{eqnarray}
\label{conditions}
[\psi]=0,~~~[\nu]=0,
\end{eqnarray}
where $ [x] $ denotes the difference between the values of $ x $ inside and outside evaluated at $r=R$, i.e., $ [x]=x^{-}|_{r=R}-x^{+}|_{r=R}.$ We use $ +~(-) $ to describe quantities in the stellar exterior (interior). For later convenience, we introduce the stellar compactness defined by
\begin{align}
    {\cal C}=\frac{M}{R}.\label{eq:compactness}
\end{align}

\subsubsection{$\mathcal{O}(\epsilon^1)$: Poloidal Magnetic Field Configuration}

We next move to $\mathcal{O}(\epsilon^1)$ where we solve for the magnetic field. We focus on the purely poloidal magnetic field, where the vector potential is given by $A_{\mu} = (0,0,0,\epsilon\,A_{\phi})$ that is sourced by a toroidal 4-current $J_{\mu}=(0,0,0,\epsilon\,J_{\phi})$. 
We expand the functions $ A_{\phi}(r,\theta) $ and $ J_{\phi}(r,\theta) $ in terms of the Legendre polynomials $ P_{l}(\cos\theta) $ as \cite{Regge:1957td}
\begin{align}
\label{eq:A-J_exp}
A_{\phi} (r,\theta) =& \sum_{l=1}^{\infty} a_{l}(r) \sin\theta \frac{dP_{l}(\cos\theta)}{d\theta}, \\
J_{\phi}(r,\theta) = & \sum_{l=1}^{\infty} j_{l}(r) \sin\theta \frac{dP_{l}(\cos\theta)}{d\theta}.
\end{align}

Henceforth, we focus on the dipole 
field, i.e., $ l=1 $, and the Maxwell equation reduces to
\begin{eqnarray}
\label{a1eqq}
e^{-\psi} \frac{d^2 a_{1}}{d r^2} + \frac{1}{2} \left(\frac{d\nu}{dr}-\frac{d\psi}{dr} \right) e^{-\psi} \frac{da_{1}}{dr} -\frac{2}{r^2} a_{1} = -4\pi j_{1} , \nonumber \\
\end{eqnarray}
where $a_1$ and $j_1$ are given in Eq.~\eqref{eq:A-J_exp}. The expression for $j_1$ is obtained at level of $\mathcal{O}(\epsilon^2)$, which we will discuss in the Appendix~\ref{app:pert_eqs}.

The boundary conditions are obtained by matching the interior and exterior solutions at the stellar surface.
In the exterior region (where $\rho=p=0$ which further leads to $j_{1}=0$ since there is no matter to carry the current), the vector potential is analytically solved in the form of~\cite{Wasserman1983} 
\begin{equation}
\label{a1ext}
a_{1}^\mathrm{ext}(r) = -\frac{3\mu}{8M^3} r^2
\left[\ln\left( 1- \frac{2M}{r}   \right) + \frac{2M}{r} + \frac{2M^2}{r^2} \right],
\end{equation}
where $ \mu $ is a magnetic dipole moment. 
The boundary conditions are then given by
\begin{eqnarray}
\label{eq:a1_BC}
[a_{1}]=0,\quad [a_{1}']=0.
\end{eqnarray}

The magnetic fields in the radial and tangential directions can be written in the tetrad components, $e^{\mu}_{\hat{a}}$, which are a set of four orthonormal basis vectors. The expression reads \cite{Konno:1999zv} 
\begin{eqnarray}
\label{magnetic}
B_{\hat{r}} = \frac{2\cos\theta}{r^2}a_{1},~~~~  \textrm{and}~~~~ B_{\hat{\theta}} = -\frac{e^{-\psi/2} \sin\theta}{r}a_{1}', \nonumber \\ 
\end{eqnarray}
where a prime denotes the derivative with respect to $r$.

\subsubsection{$\mathcal{O}(\epsilon^2)$: Deformed Spacetime Configuration due to Magnetic Fields}

We finally describe the formulation at $\mathcal{O}(\epsilon^2)$. The dipole ($l=1$) vector potential at $\mathcal{O}(\epsilon)$ induces the monopole~($l=0$) and quadrupole~($l=2$) contributions to the spacetime at the present order. The ansatz for the line element including both the background and perturbations is given by
\begin{eqnarray}
ds^2 &=& -e^{\nu(r)} [1+2\epsilon^2\,h(r,\theta)] dt^2 \nonumber \\
& &+ e^{\psi(r)} \left[ 1+ 2\epsilon^2\,\frac{\mathcal{M}(r,\theta)}{r} \right] dr^2 \nonumber \\ 
&&+ r^2 [1+2\epsilon^2\,k_{2}(r)P_{2}(\cos\theta)] (d\theta^2+\sin^2\theta d\phi^2), \nonumber \\
\end{eqnarray}
where $ h(r,\theta) $ and $ \mathcal{M}(r,\theta) $ are 
expanded in terms of the Legendre polynomials as
\begin{eqnarray}
h(r,\theta) &=& h_{0}(r) + h_{2}(r)P_{2}(\cos\theta),\\ 
\mathcal{M}(r,\theta) &=& m_{0}(r) + m_{2}(r)P_{2}(\cos\theta).
\end{eqnarray}
Here, $ (h_{0},m_{0}) $ 
correspond to spherical deformation, 
while $ (h_{2},m_{2},k_{2}) $ represent 
quadrupolar deformation. Note that, although $m_0(r)$ gives a correction to the stellar mass, its contribution to the stellar compactness is not included in the following analysis, as it affects the universal relations only at higher order in the magnetic-field strength\footnote{See the paragraph below Eq.~\eqref{eq:mu_bar-C} for more details.}  .

We first discuss the interior matter fields. The total stress-energy tensor consists of the matter and electromagnetic contributions,
\begin{eqnarray}
T^{\mu}{}_{\nu} = T^{\mu}{}_{\nu~(\textrm{mat})} +  T^{\mu}{}_{\nu~(\textrm{EM})},
\end{eqnarray}
where the expression of each term is given by
\begin{eqnarray}
T^{\mu}{}_{\nu~(\textrm{mat})} &=& (\varrho+\wp)u^{\mu}u_{\nu} + \wp~ \delta^{\mu}{}_{\nu},\\
T^{\mu}{}_{\nu~(\textrm{EM})} &=& \frac{1}{4\pi} \left(F^{\mu\lambda}F_{\nu\lambda} -\frac{1}{4} F_{\sigma\lambda}F^{\sigma\lambda} \delta^{\mu}{}_{\nu}  \right).
\end{eqnarray}
with the Faraday tensor $F_{\mu\nu} = \partial_\mu A_\nu - \partial_\nu A_\mu$. Assuming a barotropic EoS of the form $\wp = \wp(\varrho)$, the pressure and energy density can be decomposed as

\begin{eqnarray}
\wp(r,\theta)&=& p + \epsilon^2\,[\delta p_0(r)+\delta p _{2}(r) P_{2}(\cos\theta)],\\ \varrho(r,\theta) &=&  \rho + \epsilon^2\,\frac{\rho'}{p'}[\delta p _{0}(r)+\delta p _{2}(r) P_{2}(\cos\theta)].
\end{eqnarray}
Note that although the magnetic-field backreaction at ${\cal O}(\epsilon^2)$ leads to the shift of the stellar radius, we do not include it in the following analysis because it affects the universal relations only at higher order in the magnetic field strength. However, at $\mathcal{O}(\epsilon^2)$, we can find the expression of $ \delta p  $ from the conservation law of the total energy-momentum tensor, $ \nabla_{\mu}T^{\mu}{}_{\nu}=0. $ The resulting expressions for $\delta p$ at each multipole are written in terms of $h_{0}$, $h_2$ and $a_1 j_1$, along with the zeroth-order quantities.

In particular, the spherical deformation and quadrupolar contributions lead to
\begin{eqnarray}
\delta p_{0} &=& -(\rho+p)h_{0} + \frac{2}{3r^2} a_{1}j_{1}+c_{2}(\rho+p),\\
\label{dp2}
\delta p _{2} &=& -(\rho+p) h_{2} - \frac{2}{3r^2}a_{1}j_{1}, \\ \delta p_{2}' &=& -\frac{\nu'}{2} \left( \frac{\rho'}{p'} +1 \right) \delta p_{2}-(\rho+p) h_{2}' - \frac{2}{3r^2}a_{1}' j_{1}. \nonumber \\
\end{eqnarray}
It follows from the integrability condition for these last two equations that $j_1$ reads
\begin{eqnarray}
\label{current}
j_{1}=c_{0} r^2 (\rho+p),
\end{eqnarray}
for a constant $c_0$. Equation~\eqref{current} allows us to obtain $a_1$ inside the star through Eq.~\eqref{a1eqq}. The detail of the perturbed Einstein field and its matching condition with the exterior is presented in the Appendix \ref{app:pert_eqs} 

\subsection{Background spherical configurations and magnetic field profiles}

In this section, we present the mass-radius relation and the magnetic field profile by solving the equations at $\mathcal{O}(\epsilon^0)$ and $\mathcal{O}(\epsilon^1)$.  The details of the numerical procedures to solve the field equations at each order are provided in Appendix~\ref{app:numerical}. 

To close the system, we need to provide an EoS to connect pressure and energy density. We use a few different types of EoSs. The first class is ``realistic'' EoSs for hadronic matter provided in terms of tables. We use the following set in this paper:  AP3 and AP4 \cite{Akmal:1998cf}, ENG \cite{Engvik:1995gn}, MPA1 \cite{Muther:1987xaa}, MS1 and MS1b \cite{Mueller:1996pm}, SLy \cite{Douchin:2001sv}, Shen \cite{Shen:1998gq,Lim:2014sra}, WFF1 and WFF2 \cite{Wiringa:1988tp}. For reference, we also consider a polytropic EoS of the form
\begin{equation}
p=\kappa \rho^{1+\frac{1}{n}},
\end{equation}
with a constant $\kappa$ and a polytropic index $n$. In particular, we consider $n=1$ with $\kappa = 2.4 \times~10^{2} ~\textrm{km}^2$ in geometric units.

\begin{figure}[t!]
	\centering
	\includegraphics[width=0.9\linewidth]{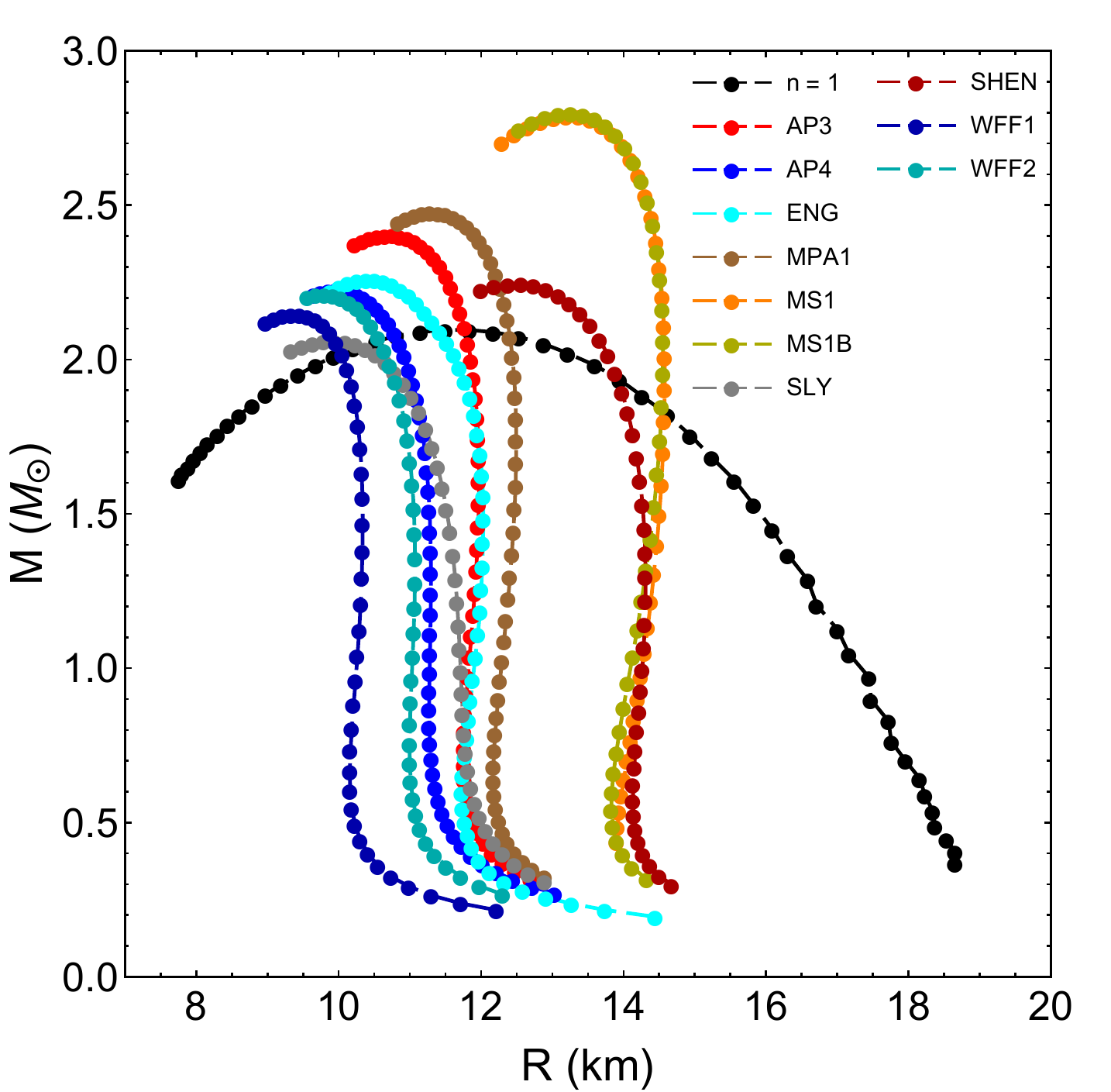}
	\caption{Mass-radius relations of background stellar configurations for the realistic EoSs considered in this work and the $n=1$ polytrope. 
    }
	\label{fig:M-R}
\end{figure}

Figure~\ref{fig:M-R} shows the mass-radius relations of background stellar configurations for the realistic EoSs \cite{Akmal:1998cf,Engvik:1995gn,Muther:1987xaa,Mueller:1996pm,Douchin:2001sv,Shen:1998gq,Lim:2014sra,Wiringa:1988tp} as well as $n=1$ polytropic EoS.
Each curve is obtained by fixing an EoS and varying the central pressure~$p_c$. 
As $p_{c}$ increases, the configurations generically transition to a dynamically unstable branch once they reach the maximum mass for each EoS.

\begin{figure*}[t!]
	\centering
	\includegraphics[width=0.45\linewidth]{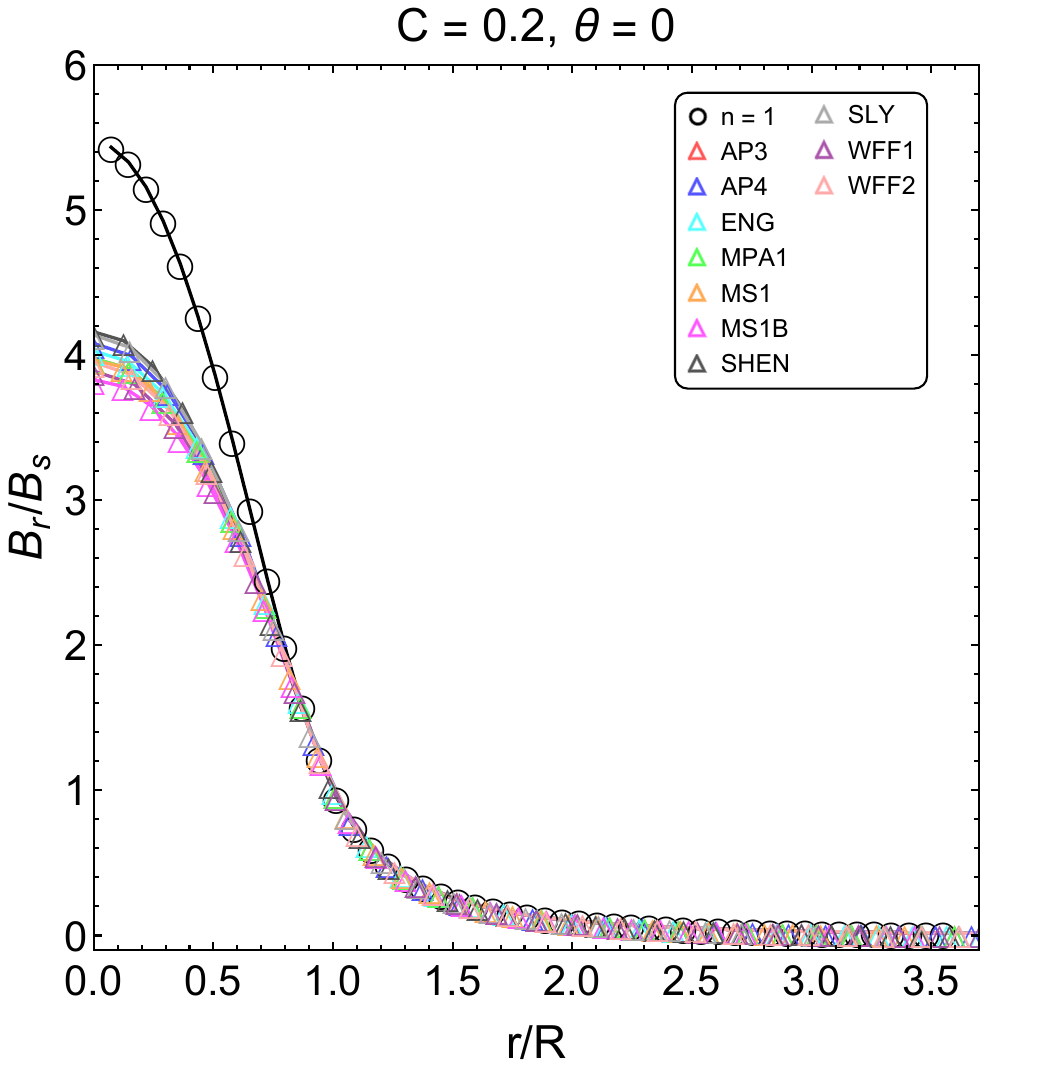}
	\includegraphics[width=0.45\linewidth]{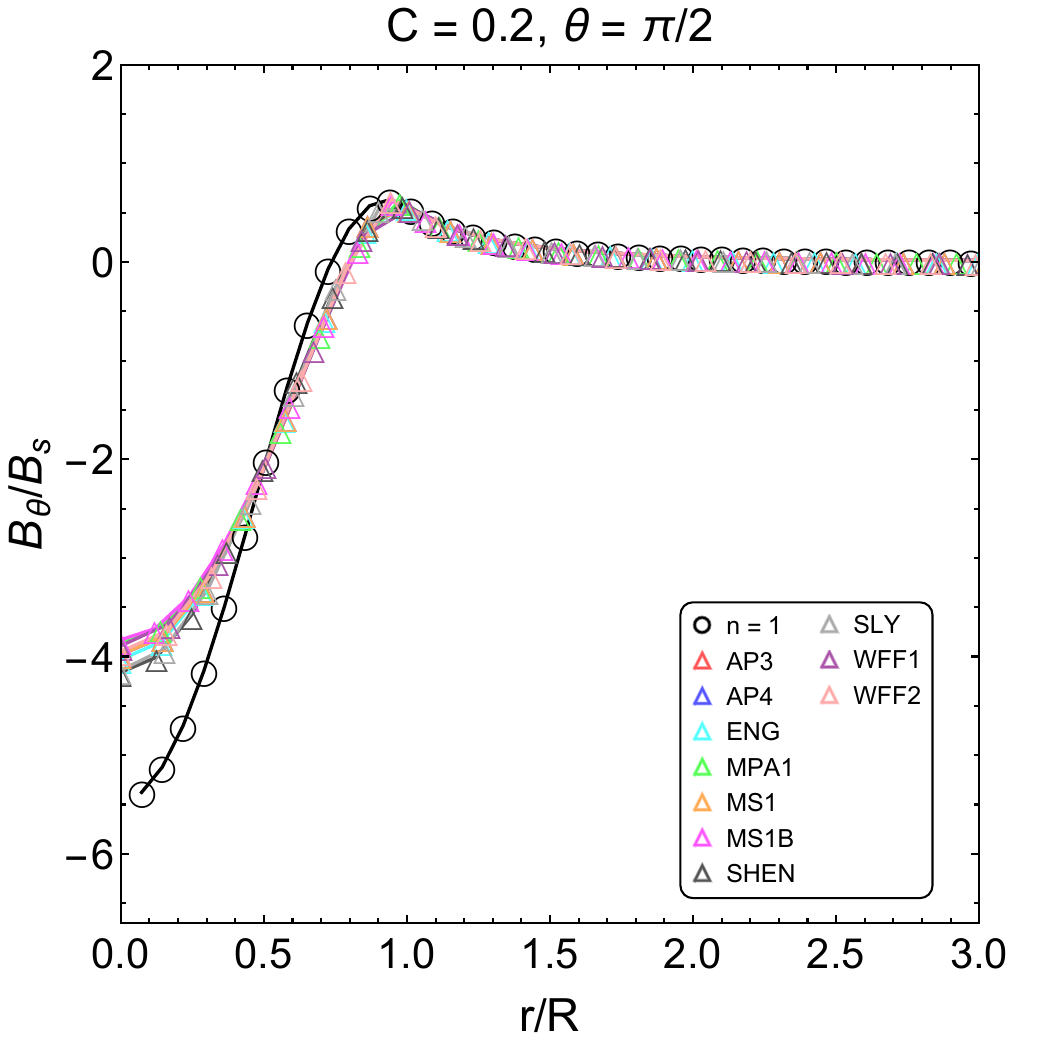}
	\caption{Magnetic field profiles for $ B_{\hat{r}} $ along the magnetic field axis $(\theta=0)$ and $ B_{\hat{\theta}} $ on the equatorial plane $(\theta=\pi/2)$,  normalized by $ B_{\textrm{s}} $, for NSs with $ \mathcal{C}=0.2$ and for various EoSs. The surface of the star is located at $r/R=1.$}
	\label{fig:magnetic}
\end{figure*}

Figure~\ref{fig:magnetic} presents the magnetic field profiles for $B_{\hat r}$ along the field axis and $B_{\hat \theta}$ on the equatorial plane for stars with compactness~$\mathcal{C}=0.2$ for various EoSs. Here, the profile is normalized by the magnetic field strength at the pole ($\theta=0$) of the surface ($r=R$), defined by
\begin{equation}
\label{eq:Bs}
    B_s = \frac{2}{R^2}a_1(R).
\end{equation}
Notice that the field outside the star is less sensitive to the choice of the EoSs, since the functional form of the exterior solution for $a_1$ in Eq.~\eqref{a1ext} is common to all EoSs (though quantities that enter in the equation like the mass $M$ depend on the EoSs for a fixed $\mathcal{C}$). Inside the star, $B_{\hat r}$ is always positive, while $B_{\hat \theta}$ is mostly negative. The latter means that $B_{\hat \theta}$ on the equatorial plane inside the star points in the opposite direction to that in the exterior region. This is typical for a poloidal field configuration. The $n=1$ magnetic profile in Fig.~\ref{fig:magnetic} is consistent with Fig.~1 of~\cite{Folomeev:2015aua}.

\section{Universal relations for magnetized neutron stars}
\label{sec:Results}

In this section, we numerically solve perturbed field equations at $\mathcal{O}(\epsilon^2)$ and present universal relations among several parameters that characterize  magnetized NSs. We also perform analytic estimates in the Newtonian limit to support our numerical findings. 

\subsection{Relevant physical parameters}

We use four dimensionless parameters to characterize universal relations. The first quantity is the background stellar compactness defined in Eq.~\eqref{eq:compactness}. In what follows, we introduce the remaining three quantities associated with the magnetic field.

The second quantity is a dimensionless magnetic dipole moment,
\begin{equation}
\label{eq:mu_bar}
    \bar \mu = \frac{\mu}{B_s M^3},
\end{equation}
where $B_s$ is the magnetic-field strength defined in Eq.~\eqref{eq:Bs}, while the magnetic dipole moment $\mu$ enters in the exterior solution for $a_1$ in Eq.~\eqref{a1ext}. The third quantity is a dimensionless magnetically-induced quadrupole moment,\footnote{A similar dimensionless quadrupole moment is defined for the spin-induced one in~\cite{Yagi:2013bca,Yagi:2013awa}} 

\begin{eqnarray}
\bar{Q} = -\frac{Q}{B_{s}^{2}M^{5}}.
\end{eqnarray}
Here, $Q$ is the magnetically-induced quadrupole moment,
\begin{equation}
Q = - \frac{8}{5}KM^{3} + \frac{6\mu^2}{5M},
\end{equation}
which is extracted from the coefficient of the term proportional to $1/r^3$ in the large-distance expansion of $h_2$ in Eq.~\eqref{h2ext} ($K$ is an integration constant). The last quantity is a dimensionless stellar ellipticity,
\begin{equation}
\bar{\varepsilon} \equiv \frac{\varepsilon}{B_{s}^2M^2}.
\end{equation}
In general, the stellar ellipticity, $\varepsilon$, 
measures the stellar shape's quadrupolar deformation and is given by~\cite{Konno:1999zv,Konno:2000tw}

\begin{equation}
\varepsilon=\frac{\textrm{(equatorial radius)}-\textrm{(polar radius)}}{\textrm{(mean radius)}}.
\end{equation}
 The ellipticity for magnetized NSs, $\varepsilon$, reads ~\cite{Konno:1999zv}
 
\begin{equation}
\label{eq:ellip}
\varepsilon = \frac{2c_{0}a_{1}}{r\nu'} + \frac{3h_{2}}{r\nu'} - \frac{3}{2}k_{2}\bigg|_{r=R},
\end{equation}
where $c_0$ enters in the dipole current $j_1$ in Eq.~\eqref{current}. The first term is due to the Lorentz force, the second term is a perturbation factor of the gravitational part induced by the magnetic field, and the last term is purely relativistic \cite{Konno:1999zv}.

\subsection{Novel universality}
\label{sec:univ_num}
In what follows, we present our main results of this work about the novel (approximate) universality for magnetized NSs.

\begin{figure}[t!]
	\centering
	\includegraphics[width=1\linewidth]{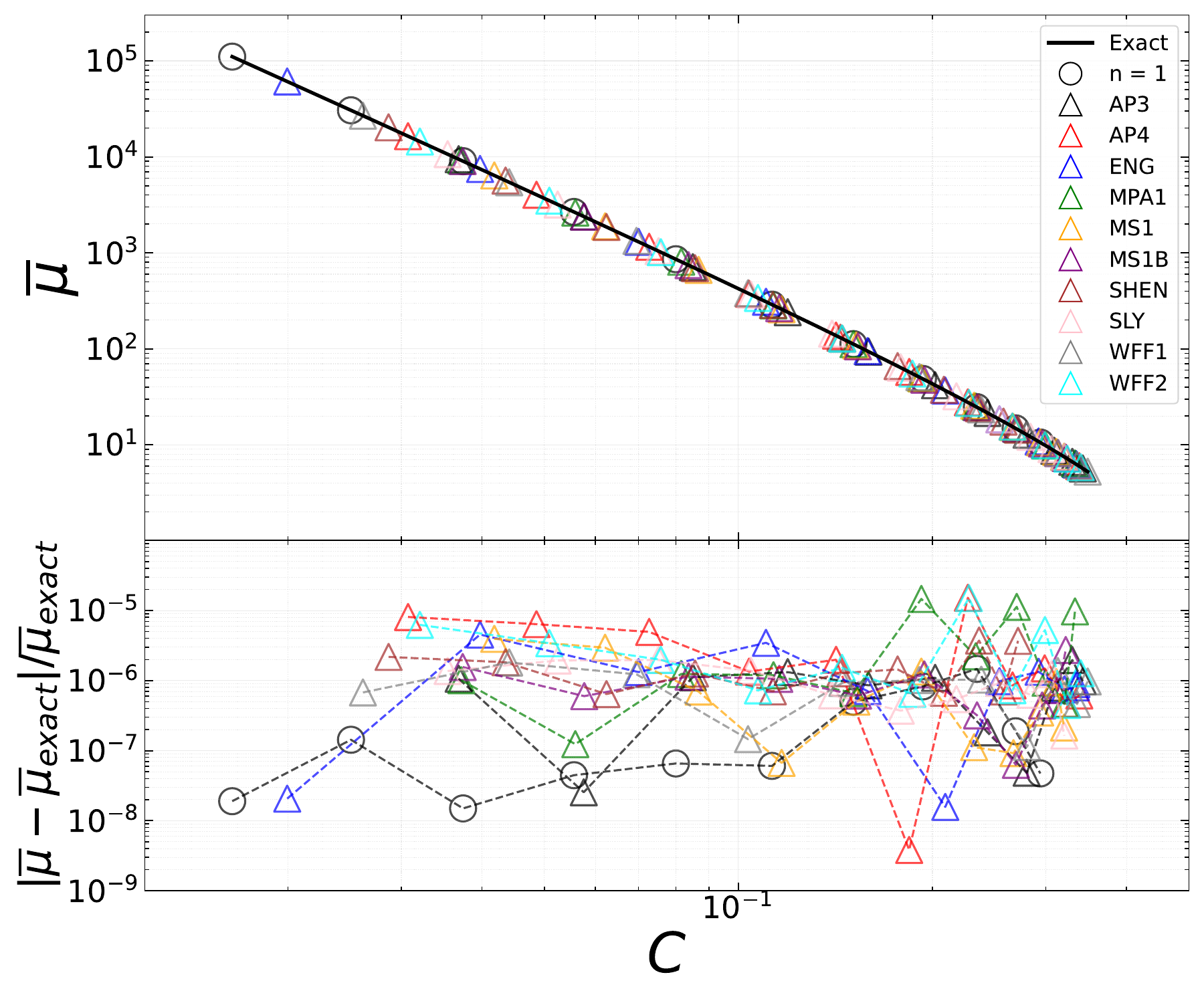}
	\caption{ (Top) Relation between the dimensionless dipole magnetic moment ($\bar{\mu}$) and the compactness ($\mathcal{C}$) for various EoSs.(Bottom) The fractional difference between the numerical data and the exact relation~\eqref{eq:mu_bar-C}. This comparison is shown to demonstrate the accuracy of the numerical calculations.}
	\label{fig:miuc}
\end{figure}

Figure~\ref{fig:miuc} presents the universal relation between $\bar{\mu}$ and $\mathcal{C}$ for various EoSs. It is worth emphasizing that this relation is exact and is independent of the internal structure of a NS (namely EoSs). Indeed, from Eqs.~\eqref{a1ext},~\eqref{eq:Bs}, and~\eqref{eq:mu_bar}, we analytically find the unique relation,
\begin{equation}
\label{eq:mu_bar-C}
    \bar \mu = -\frac{4}{3 [2 \mathcal{C} (\mathcal{C}+1)+\log (1-2 \mathcal{C})]}.
\end{equation}
Notice that the results in Fig.~\ref{fig:miuc} are in good agreement with the exact relation in Eq.~\eqref{eq:mu_bar-C}. The bottom panel of Fig.~\ref{fig:miuc} thus shows the accuracy of our numerical calculation. 

We note that Eq.~\eqref{eq:mu_bar-C} is valid only at leading order in the perturbative expansion, namely $\mathcal{O}(\epsilon^0)$. To incorporate the next-to-leading-order correction at $\mathcal{O}(\epsilon^2)$ in the universal relation, it is not sufficient to determine the compactness to $\mathcal{O}(\epsilon^2)$ alone. One must also compute the magnetic dipole moment to $\mathcal{O}(\epsilon^3)$, since this contributes at $\mathcal{O}(\epsilon^2)$ to the normalized magnetic dipole moment $\bar{\mu}$. Such a calculation is beyond the scope of the present paper.

\begin{figure*}[t!]
	\centering
	\includegraphics[width=0.49\linewidth]{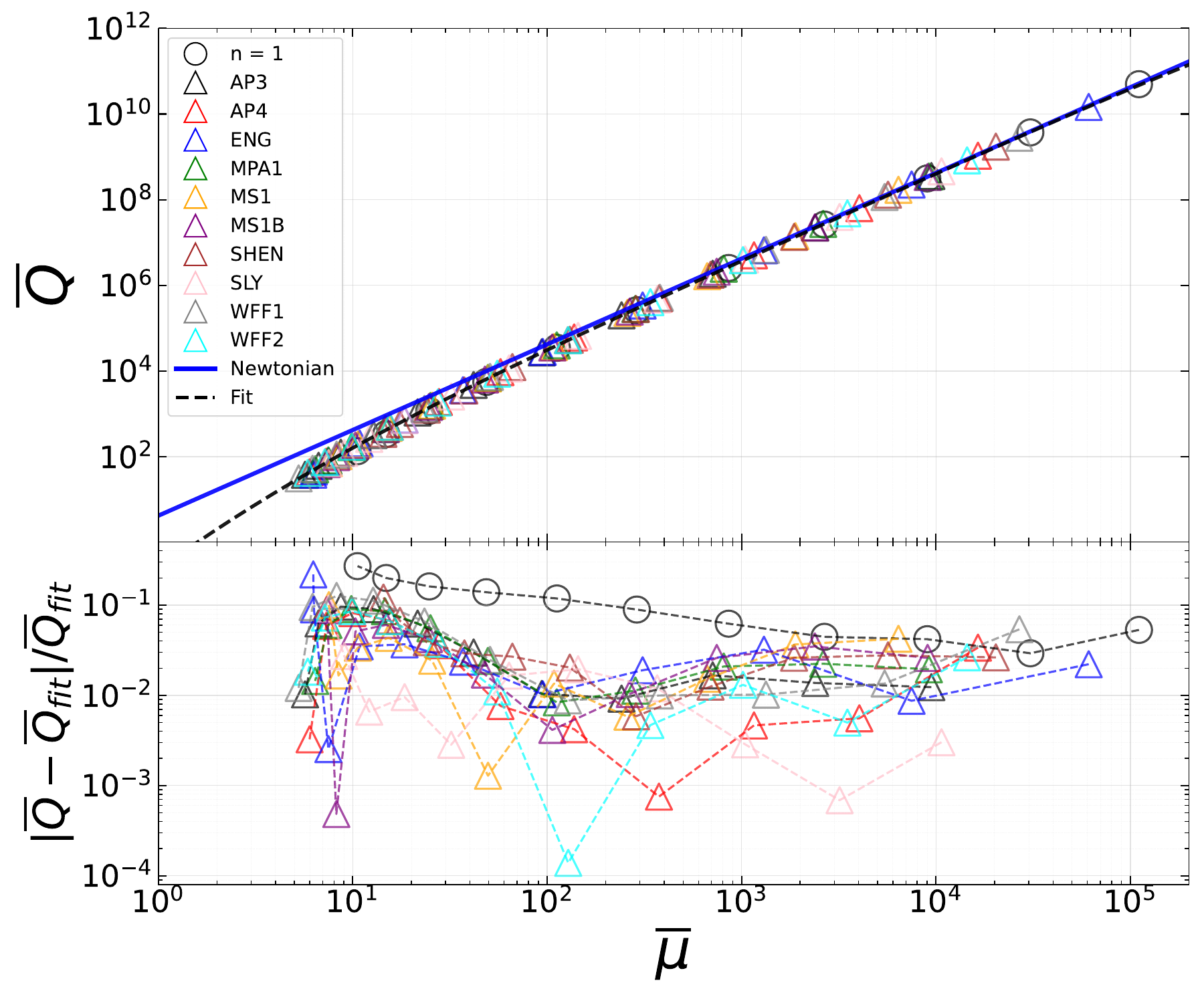}
	\includegraphics[width=0.49\linewidth]{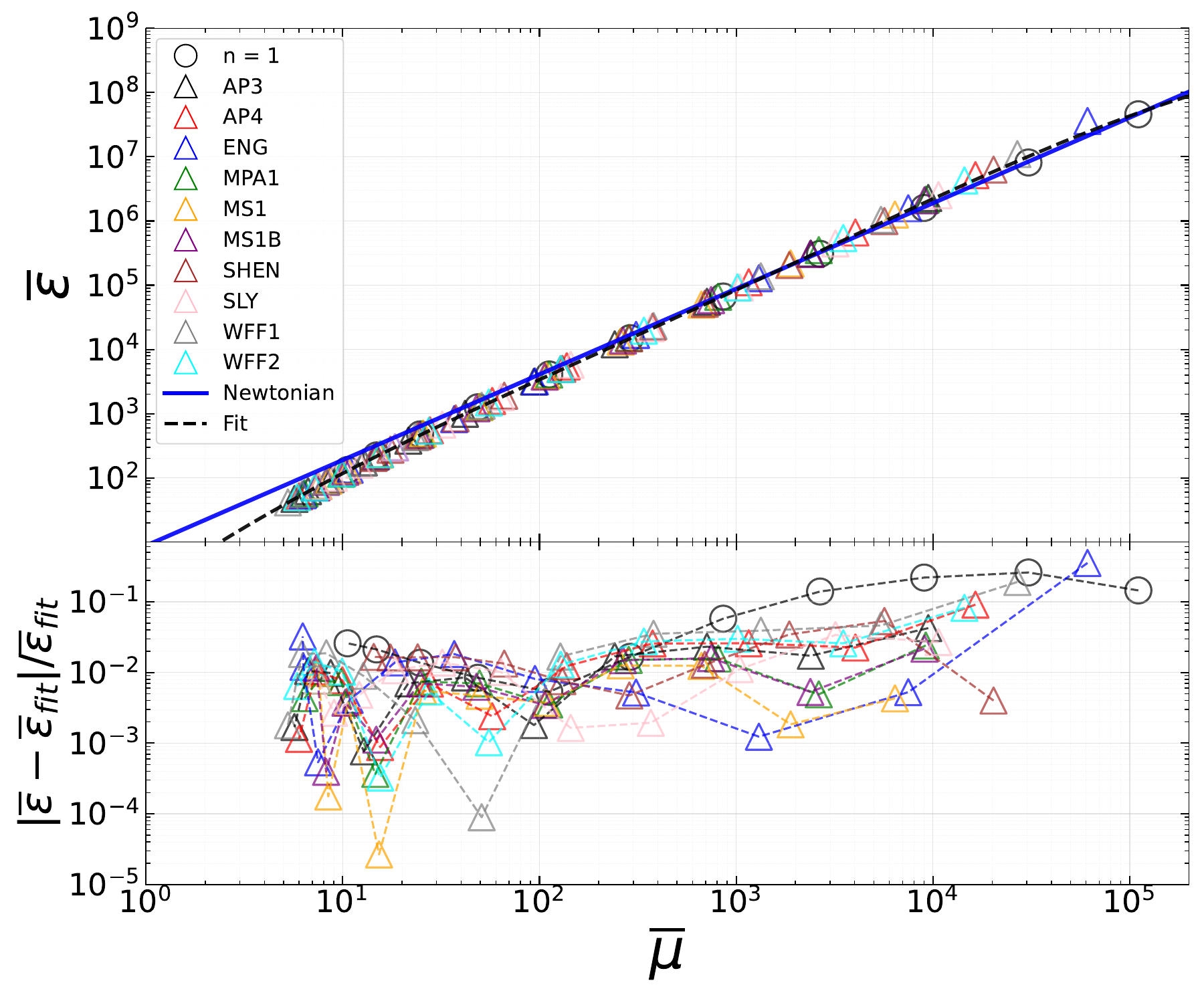}
	\includegraphics[width=0.49\linewidth]{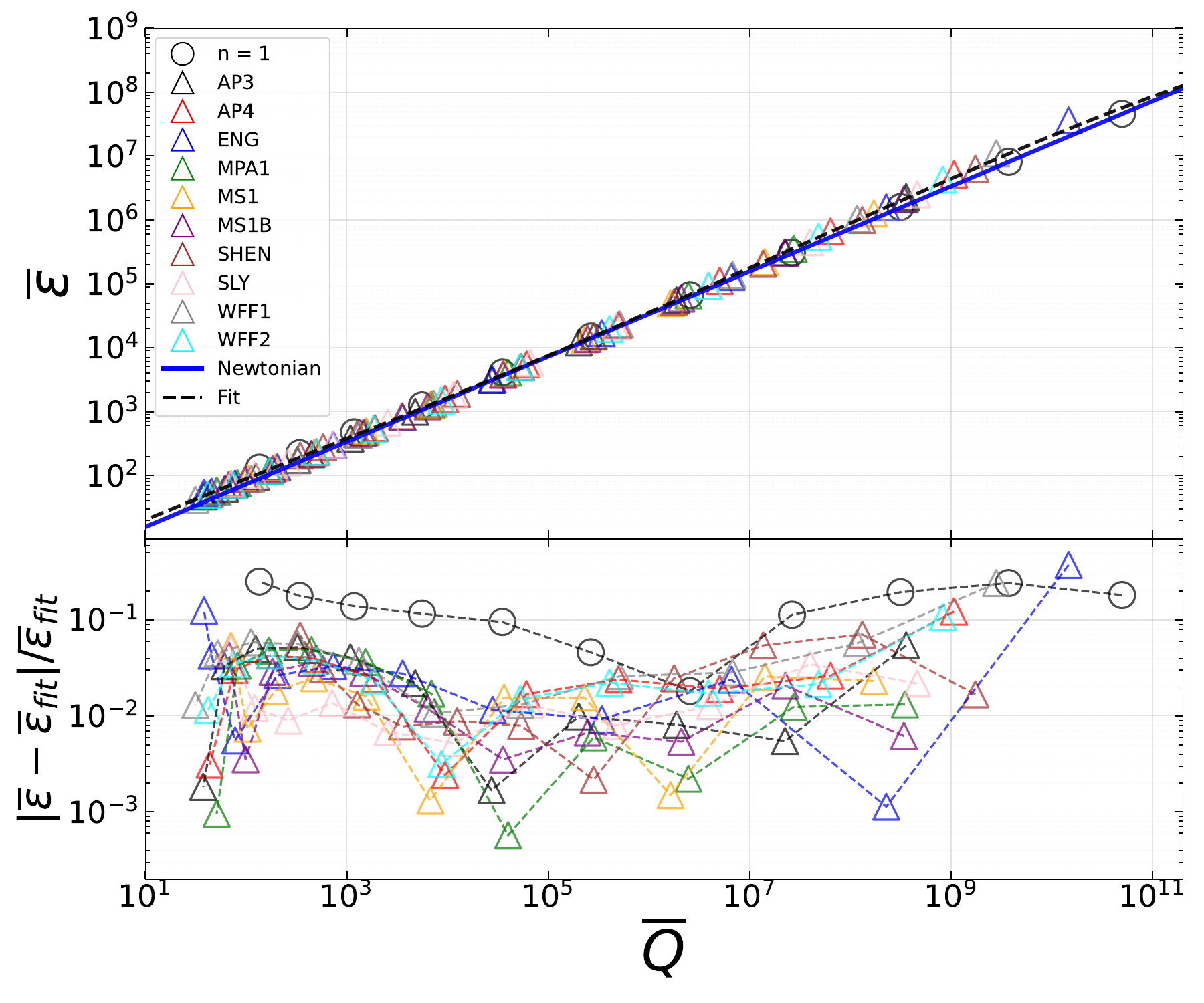}
	\caption{
    Similar to Fig.~\ref{fig:miuc} but for the $\bar Q$--$\bar \mu$ (top left), $\bar \varepsilon$--$\bar \mu$ (top right), and $\bar \varepsilon$--$\bar Q$ (bottom) relations. We also present the Newtonian relation for $n=1$ polytropes (derived in Sec.~\ref{sec:analytic}) and the fit in Eq.~\eqref{fiteq}. The bottom part of each panel shows the fractional difference of numerical data points from the fit.}
    \label{fig:univ}
\end{figure*}

Figure~\ref{fig:univ} presents approximate universal relations involving $\bar Q$, $\bar \varepsilon$, or $\bar \mu$. For each relation, we construct a fitting function in the form
\begin{eqnarray}
\label{fiteq}
	\ln~y_{i} &=& a_{i} + b_{i}~\ln x_{i} +c_{i}~ (\ln~x_{i})^2 + d_{i}~ (\ln~x_{i})^3 \nonumber \\
    &&+e_{i}(\ln x_{i})^4,
\end{eqnarray}
and provide the fitting coefficients in Table~\ref{table_fit}. 
In the bottom half of each panel, we provide fractional errors between each numerical data and the fit, demonstrating the EoS variation in each relation. It is worth noting that, for realistic EoSs, the EoS variation is $\lesssim 5\%$ in a large range of compactness. The error increases to $\mathcal{O}(10\%)$ or larger in the small $\bar \mu$ regime for the $\bar Q$--$\bar \mu$ relation and in the large $\bar \mu$ or $\bar Q$ regime for the $\bar \varepsilon$--$\bar \mu$ or $\bar \varepsilon$--$\bar Q$ relation.  Notice that one can easily convert the relations involving $\bar \mu$ to those with $\mathcal{C}$ through the one-to-one correspondence in Eq.~\eqref{eq:mu_bar-C}.

The $n=1$ polytrope exhibits the largest deviations from the universal relations among the EoSs considered, with EoS-induced variations reaching up to $\mathcal{O}(10\%)$. Unlike realistic neutron-star EoSs, which satisfy common microphysical and astrophysical constraints and therefore produce broadly similar stellar structures, the $n=1$ polytrope is a simplified toy model with a markedly different internal density profile. These differences are reflected in both the mass-radius relation shown in Fig.~\ref{fig:M-R} and the magnetic-field profile shown in Fig.~\ref{fig:magnetic}. The larger departures of the $n=1$ polytrope from the universal relations are therefore consistent with its less realistic internal structure, while the closer agreement among realistic EoSs suggests that their shared structural properties contribute to the observed universality.

\begin{table*}[t]
\label{table1}
\centering
 \begin{tabular}{||c c c c c c c||} 
 \hline
 $y_{i}$ & $x_{i}$ & $a_{i}$ & $b_{i}$ & $c_{i}$ & $d_{i}$ & $e_{i}$ \\ [0.5ex] 
 \hline \hline
 
  &  &  &  &  &  &  \\
 $\bar{Q}$ & $\bar{\mu}$ & $-1.45$ & $3.27$ & $-2.3\times10^{-1}$ & $1.92\times10^{-2}$ & $-6.09\times10^{-4}$ \\ [2ex]
 $\bar{\varepsilon}$ &  $\bar{\mu}$ & $5.35\times10^{-1}$ & $2.19$ & $-1.97\times10^{-1}$ & $2.09\times10^{-2}$ & $-7.94\times10^{-4}$ \\ [2ex] $\bar{\varepsilon}$ &  $\bar{Q}$ & $1.48$ & $6.86\times10^{-1}$ & $-1.14\times10^{-2}$ & $9.23\times10^{-4}$  & $-2.01\times10^{-5}$ \\ [2ex] 
 \hline
 \end{tabular}
 \caption{
 \label{table_fit}
 Coefficients of numerical fitting for, $\bar{Q}$-$\bar{\mu}$, $\bar{\varepsilon}$-$\bar{\mu}$, and $\bar{Q}$-$\bar{\varepsilon}$ relations given by \eqref{fiteq}}.
\end{table*}

\subsection{Analytic study}
\label{sec:analytic}
We provide analytic estimates of the universal relations presented in the previous section by working in the Newtonian limit ($m \ll r$ and $p \ll \rho$). 

\subsubsection{Perturbation equations and exterior solutions}
We reduce the field equations derived in Sec.~\ref{sec:MagnetizedNSs} to the Newtonian counterparts at each order. 

At ${\cal O} (\epsilon)$, Eq.~\eqref{a1eqq} in the Newtonian limit reduces to
\begin{eqnarray}
\label{eq:a1eq_N}
    a_{1,N}''-\frac{2 a_{1,N}}{r^2}=-4 \pi  c_0 \rho r^2,
\end{eqnarray}
where the subscript N refers to the Newtonian limit.
The exterior solution for $a_{1,N}$ in the above equation becomes
\begin{eqnarray}
\label{eq:a1_N_ext}
    a_{1,N} = \frac{\mu}{r} = \frac{B_s R^3}{2r},
\end{eqnarray}
where we have used 
$B_s$ in Eq.~\eqref{eq:Bs} to rewrite $\mu$ in terms of this quantity and the stellar radius.

At $\mathcal{O}(\epsilon^2)$, 
Eqs.~\eqref{nu2eq} and~\eqref{h2eq} reduce to
\begin{widetext}
\begin{eqnarray}
\label{eq:v2_eq_N}
    v_{2,N}'+\frac{2 m
  h_{2,N}}{r^2} &=&\frac{4 a_{1,N} a_{1,N}'}{3 r^2}+\frac{2 a_{1,N}'^2}{3 r}, \\
   \label{eq:h2_eq_N}
   h_{2,N}'-
   \left(\frac{4 \pi  r^2 \rho}{m}-\frac{2}{r}\right)h_{2,N}+\frac{2
   v_{2,N}}{m}&=&\frac{a_{1,N}'^2}{3 m}+\frac{4 a_{1,N} a_{1,N}'}{3 m r}+\frac{8 \pi 
   c_0 r^2 a_{1,N} \rho}{3 m}+\frac{4 a_{1,N}^2}{3 m r^2}
\end{eqnarray}
\end{widetext}
where we used the Newtonian relations
\begin{eqnarray}
    \nu' = \frac{2m}{r^2}, \quad e^{\psi} = 1 + \frac{2m}{r}. 
\end{eqnarray}
The exterior solutions are given by
\begin{equation}
\label{eq:h2_v2_ext_N}
    h_{2,N}^\mathrm{ext} = \frac{Q_{N}}{r^3}, \quad v_{2,N}^\mathrm{ext} = \frac{\mu_{N}^2+3 M Q_{N}}{6 r^4}.
\end{equation}

For the stellar ellipticity, we can ignore the last term in Eq.~\eqref{eq:ellip}, as it is a purely relativistic effect. Using the exterior solutions for $a_{1,N}$ and $h_{2,N}$ above, we find the normalized stellar ellipticity as
\begin{eqnarray}
\label{eq:ellip_N}
    \bar \varepsilon = \frac{c_0 \bar \mu_N}{B_s}+\frac{3 \mathcal{C}^2 \bar Q_{N}}{2}.
\end{eqnarray}

Below, we estimate $\bar Q_{N}$ and $\bar \varepsilon_{N}$ for incompressible ($n=0$) and $n=1$ polytropes.

\subsubsection{Incompressible Stars}
At $\mathcal{O}(\epsilon^0)$, the energy density and mass function inside an incompressible star are given by
\begin{eqnarray}
    \rho^\mathrm{(n=0)} = \rho_c, \quad m^\mathrm{(n=0)} = \frac{4\pi}{3} \rho_c r^3.
\end{eqnarray}
At $\mathcal{O}(\epsilon)$, the solution to 
Eq.~\eqref{eq:a1eq_N} (that is regular at $r=0$) becomes
\begin{eqnarray}
    a_{1,N}^\mathrm{(n=0)}  = a_c r^2-\frac{3 c_0 M }{10 R^3}r^4.
\end{eqnarray}
Matching this interior solution with the exterior solution in Eq.~\eqref{eq:a1_N_ext} at the surface, one can determine $a_c$ and $c_0$ in terms of $B_s$. One can then rewrite the vector potential function as
\begin{eqnarray}
      a_{1,N}^\mathrm{(n=0)} = \frac{1}{4} B_s r^2 \left(5-\frac{3 r^2}{R^2}\right).
\end{eqnarray}

Let us next estimate $\bar Q_B$ and $\bar \varepsilon_B$.
At $\mathcal{O}(\epsilon^2)$, the solutions to the equations for $h_{2,N}$ and $v_{2,N}$ in Eqs.~\eqref{eq:v2_eq_N} and~\eqref{eq:h2_eq_N} can be found as
\begin{eqnarray}
    h_{2,N}^\mathrm{(n=0)} &=& A r^2,\\
    v_{2,N}^\mathrm{(n=0)} &=& -2A\,  \mathcal{C}
   \frac{r^4}{R^4}+\frac{B_s^2 r^2}{24}\left(100 -105 \frac{r^2}{R^2}+36 \frac{r^4}{R^4}
    \right). \nonumber \\
\end{eqnarray}
Matching these interior solutions with the exterior solutions in Eqs.~\eqref{eq:h2_v2_ext_N}
 at the surface, one can determine $A$ and $Q_B$ in terms of $B_s$. One then finds
\begin{eqnarray}
      h_{2,N}^\mathrm{(n=0)} &=& \frac{5 B_s^2 }{4\, \mathcal C}r^2,\\
    v_{2,N}^\mathrm{(n=0)} &=& \frac{B_s^2}{6}r^2 \left(5-3 \frac{r^2}{R^2}\right)^2.
\end{eqnarray}
The normalized $Q$ is given by
 \begin{equation}
 \label{qmu_n0}
     \bar Q^\mathrm{(n=0)} = \frac{5}{4\, \mathcal{C}^6} = 5 \bar \mu^2,
 \end{equation}
 where we used the Newtonian limit of Eq.~\eqref{eq:mu_bar-C} given by $\bar \mu = \mathcal{C}^{-3}/2$. The normalized ellipticity, on the other hand, can be found from Eq.~\eqref{eq:ellip_N} as
 \begin{eqnarray}
 \label{eq:ellip_N_n0}
    \bar \varepsilon^\mathrm{(n=0)} =\frac{25}{2^{5/3}} \bar \mu_N^{4/3} 
    = \frac{5^{4/3}}{2^{5/3}}\bar Q^{2/3}. 
 \end{eqnarray}
 
\subsubsection{$n=1$ Polytropes}

At $\mathcal{O}(\epsilon^0)$, the energy density for an $n=1$ polytrope can be obtained from the Lane-Emden solution as
\begin{equation}
    \rho^\mathrm{(n=1)} = \frac{M}{4R^2 r}\sin \left(\frac{\pi  r}{R}\right).
\end{equation}
The mass function is then given by
\begin{eqnarray}
   m^\mathrm{(n=1)} = \frac{M}{\pi }\sin \left(\frac{\pi  r}{R}\right)-\frac{M}{R} r \cos
   \left(\frac{\pi  r}{R}\right).
\end{eqnarray}
At $\mathcal{O}(\epsilon)$, one can solve for the vector potential in the interior region, which is then matched with the exterior one at the surface to yield
\begin{eqnarray}
      a_{1,N}^\mathrm{(n=1)} &=& \frac{B_s R^3}{2 \pi  \left(\pi ^2-6\right) r} \left[\pi ^3 \frac{r^3}{R^3}-3 \left(2-\pi ^2 \frac{r^2}{R^2}\right) \sin
   \left(\frac{\pi  r}{R}\right) \right. \nonumber \\
   && \left. +6 \pi  \frac{r}{R}\cos \left(\frac{\pi 
   r}{R}\right)\right].
\end{eqnarray}

Let us now estimate $\bar Q_B$ and $\bar \varepsilon_B$. At $\mathcal{O}(\epsilon^2)$, $h_{2,N}$ and $v_{2,N}$ can be solved in the interior region, which are once again matched with the exterior solutions at the surface.  For example, $h_{2,N}$ is given by
\begin{eqnarray}
      h_{2,N}^\mathrm{(n=1)} &=& \frac{\pi B_s^2 R^5}{4
   \left(\pi ^2-6\right)^2 \mathcal{C} r^3} \nonumber \\
   && \times \left\{3 \left[12-\pi ^2  \frac{r^2}{R^2}  \left(\frac{r^2}{R^2}+4
    \right)\right] \sin \left(\frac{\pi  r}{R}\right) \right. \nonumber \\
    && \left.-\pi  \frac{r}{R} \left(36-\pi ^2 \frac{r^4}{R^4}
   \right) \cos \left(\frac{\pi  r}{R}\right)-2 \pi ^3 \frac{r^5}{R^5}\right\}. \nonumber \\
\end{eqnarray}
Notice that this functional form is same as the magnetic field correction to the gravitational potential for $n=1$ polytropes in~\cite{Haskell:2007bh}. 
A similar but slightly longer expression is found for $v_{2,N}$.
$\bar Q_{N}^\mathrm{(n=1)}$ is given by
\begin{eqnarray}
\label{qmu_n1}
    \bar Q_{N}^\mathrm{(n=1)} = \frac{3 \pi ^2 \left(12-\pi ^2\right)}{4 \left(\pi ^2-6\right)^2
   \mathcal C^6} = \frac{3 \pi ^2 \left(12-\pi ^2\right)}{\left(\pi ^2-6\right)^2} \bar \mu_N^2,
\end{eqnarray}
while $\bar \varepsilon^\mathrm{(n=1)}$ is found to be
\begin{eqnarray}
\label{eq:ellip_N_n1}
    \bar \varepsilon_{N}^\mathrm{(n=1)} &=& \frac{3 \pi ^2 \left(24-\pi ^2\right)}{2^{5/3}
   \left(\pi ^2-6\right)^2}\bar\mu_N^{4/3} \nonumber \\
    &=& \frac{3^{1/3} \pi ^{2/3} \left(24-\pi ^2\right)}{2^{5/3}
   \left[\left(12-\pi ^2\right) \left(\pi ^2-6\right)\right]^{2/3}}\bar Q_{N}^{2/3}.
\end{eqnarray}

\subsubsection{Universality}
Having the above analytic results, we can compare them with the numerical results presented in Sec.~\ref{sec:univ_num}, as well as estimate the amount of EoS variation between these polytropic NSs in the universal relations. In Fig.~\ref{fig:univ}, we present the analytic relations in the Newtonian limit.  In particular, we used the Newtonian relation \eqref{qmu_n1} in the top left panel while \eqref{eq:ellip_N_n1} in the top right and bottom panels. Observe that the numerical relations correctly and accurately approach these analytic ones in the Newtonian regime (large $\bar \mu$, $\bar Q$, and $\bar \varepsilon$). Next, we take the fractional difference in the above analytic relations for Newtonian polytropes and find the following EoS variation:
\begin{widetext}
\begin{eqnarray}
    \frac{\left|\bar Q_{N}^\mathrm{(n=0)}(\bar \mu_N)-\bar Q_{N}^\mathrm{(n=1)}(\bar \mu_N)\right|}{\left( \bar Q_{N}^\mathrm{(n=0)}(\bar \mu_N)+\bar Q_{N}^\mathrm{(n=1)}(\bar \mu_N)\right)/2} &=& \frac{180-96 \pi ^2+8 \pi ^4}{90-12 \pi ^2+\pi ^4} = 0.171, \\
    \frac{\left|\bar \varepsilon_{N}^\mathrm{(n=0)}(\bar \mu_N)-\bar \varepsilon_{N}^\mathrm{(n=1)}(\bar \mu_N)\right|}{\left(\bar \varepsilon_{N}^\mathrm{(n=0)}(\bar \mu_N)+\bar \varepsilon_{N}^\mathrm{(n=1)}(\bar \mu_N) \right)/2} &=& \frac{4 \left(93 \pi ^2-225-7 \pi ^4\right)}{450-114 \pi ^2+11 \pi ^4} = 0.111, \\
     \frac{\left|\bar \varepsilon_{N}^\mathrm{(n=0)}(\bar Q_{N})-\bar \varepsilon_{N}^\mathrm{(n=1)}(\bar Q_{N})\right|}{\left(\bar \varepsilon_{N}^\mathrm{(n=0)}(\bar Q_{N})+\bar \varepsilon_{N}^\mathrm{(n=1)}(\bar Q_{N}) \right)/2} &=& 2-\frac{4\times 5^{4/3} \left[\left(12-\pi ^2\right) \left(\pi
   ^2-6\right)\right]^{2/3}}{5^{4/3}\left[\left(12-\pi ^2\right)
   \left(\pi ^2-6\right)\right]^{2/3}+3^{1/3}\pi ^{2/3} \left(24-\pi
   ^2\right)} \nonumber \\
   &=& 0.225.
\end{eqnarray}
\end{widetext}
Thus, we see that these relations are universal within an error of about 10\% to 20\%, which is roughly in agreement with the numerical results.

\section{Conclusions}
\label{sec:summary}

We modeled static, magnetized NSs by treating the effect of the magnetic field perturbatively and studied universal relations among the stellar compactness $\mathcal{C}$, the magnetic dipole moment $\bar \mu$, the magnetically-induced quadrupole moment $\bar Q$, and ellipticity $\bar \varepsilon$. The latter three quantities are normalized by the stellar mass and the magnetic field strength at the stellar pole. We first discovered an exact relation between the normalized dipole moment and compactness whose EoS variation vanishes identically. We then investigated the relations among $\bar \mu$, $\bar Q$, and $\bar \varepsilon$, finding them to be EoS-insensitive at the level of $\mathcal{O}(10\%)$. We also carried out analytic calculations for $n=0$ and $n=1$ polytropes in the Newtonian limit. We confirmed that the numerical results for $n=1$ polytropes correctly approach the corresponding analytic relation in the small-compactness regime. Furthermore, by comparing the analytic results for $n=0$ and $n=1$, we verified that the relations among $\bar \mu$, $\bar Q$, and $\bar \varepsilon$ remain (quasi-) universal, with EOS variations of only $\mathcal{O}(10\%)$, in agreement with our numerical findings.

Several avenues for future work remain. One important direction is to investigate how the universal relations discovered here depend on the magnetic-field configuration. A purely poloidal magnetic-field configuration is known to be dynamically unstable on an Alfv\'en timescale~\cite{10.1093/mnras/163.1.77,Ciolfi:2011xa,Kiuchi:2011yt,Lasky:2011un,Ciolfi:2012en}. It is therefore more realistic to consider configurations containing both poloidal and toroidal components, such as twisted-torus configurations~\cite{2009MNRAS.397..913C,2010MNRAS.406.2540C,2013MNRAS.435L..43C}. It would also be interesting to examine how these universal relations are modified in superconducting NSs~\cite{Das:2025fws}.

Another important avenue is to study the applicability of these universal relations to observations and their impact on parameter inference. For example, $\bar \mu$~\cite{Ioka:2000yb} and $\bar Q$~\cite{poisson-quadrupole}\footnote{Spin-induced quadrupole moment and magnetically-induced quadrupole moment enter at the same post-Newtonian order in the waveform phase (see footnote~\ref{footnote} for more details).} both enter the gravitational-wave phase at the same post-Newtonian order at leading order for binaries containing magnetized NSs. The universal relation between these two quantities can therefore be used to break the degeneracy between them. It would be interesting to carry out a parameter-estimation study to quantify the statistical uncertainty in the magnetic-field strength inferred from gravitational-wave observations, as well as to estimate systematic errors arising from the EOS variation in the approximate universal relation.


\acknowledgments
B.N.J. thanks Fahmi Fauzi, Ilham Prasetyo, and Haris Yunefi for the useful discussions. B.N.J. is supported by the Second Century Fund (C2F), Chulalongkorn University, Thailand and Conducting Research Abroad, Chulalongkorn University, Thailand.
T.K.\ is supported by the MUR FIS2 Advanced Grant ET-NOW (CUP:~B53C25001080001) and by the INFN TEONGRAV initiative.
S.A. and K.Y. acknowledge support from NSF Grant PHYS-2339969. K.Y. also acknowledges support from NSF Grant PHY-2309066.


\appendix

\section{Perturbation Equations and Boundary Conditions}
\label{app:pert_eqs}

In this appendix, we present perturbation equations and boundary conditions at second order in the $\epsilon$ expansion. We closely follow the formulation in~\cite{Konno:1999zv}.



\subsection{$\mathcal{O}(\epsilon^2)$: Deformed Spacetime Configuration due to Magnetic Fields}
\label{app:epsilon2}

Having both spacetime and matter perturbations in hand, the relevant components of the perturbed Einstein equations, $ \delta G^{\mu}{}_{\nu} = 8\pi \delta T^{\mu}{}_{\nu}$, reduce to \cite{Konno:1999zv}

\begin{widetext}
\begin{eqnarray}
m_{0}' &=& -4\pi r^2 \frac{\rho'}{p'}(\rho+p)(h_{0}-c_{2}) + \frac{1}{3}e^{-\lambda}(a_{1}')^2 + \frac{2}{3r^2}a_{1}^2 +\frac{8\pi}{3} \frac{\rho'}{p'}a_{1}j_{1},    \\ h_{0}' &=& \left( \frac{1}{r^2} + \frac{\nu'}{r}\right) e^{\lambda} m_{0} -4\pi r e^{\psi} (\rho+p)(h_{0}-c_{2}) + \frac{(a_{1}')^2}{3r} -\frac{2}{3r^3}e^{\psi}a_{1}^2 + \frac{8\pi}{3r}e^{\psi} a_{1} j_{1},
\end{eqnarray}
and
\begin{eqnarray}
\label{nu2eq}
v_{2}' &=& -\nu' h_{2} + \frac{2}{3} e^{-\psi} \left(\frac{1}{r}+\frac{\nu'}{2}\right)(a_{1}')^2 + \frac{4}{3r^2}a_{1}a_{1}', \\ \label{h2eq}   h_{2}' &=& -\frac{4e^{\psi}}{r^2\nu'}v_{2}+ \left[\frac{8\pi e^{\psi}}{\nu'}(\rho+p) + \frac{2}{r^2\nu'}(1-e^{\psi})-\nu' \right]h_{2}\nonumber \\&& +\frac{8}{3r^4\nu'} e^{\psi} a_{1}^2+\frac{8}{3r^3 \nu'} \left( 1+\frac{r\nu'}{2} \right) a_{1}a_{1}'+\left( \frac{1}{3}\nu' e^{-\psi} + \frac{2}{3r^2\nu'}   \right) (a_{1}')^2 + \frac{16\pi}{3r^2\nu'} e^{\psi} a_{1} j_{1}, \nonumber \\
\end{eqnarray}
\end{widetext}
where $c_{1}$ is the integration constant and $ v_{2}\equiv h_{2}+ k_{2}$.

To find the above expressions, we already substituted Eq.~\eqref{dp2} as well as the algebraic relation for $m_2$ given by 
\begin{eqnarray}
    m_2 =\frac{r}{e^{\lambda}} \left[\frac{2}{3}e^{-\psi} (a_{1}')^2 -h_{2}\right]
\end{eqnarray}
As pointed out in \cite{Ioka:2003nh}, it is convenient to introduce a new variable defined by
\begin{eqnarray}
\label{y2def}
y_{2}\equiv v_{2} -\frac{e^{-\psi}}{6} a_{1}'^{2}-\frac{2e^{-\psi}}{3r}a_{1}a_{1}' -\frac{2}{3r^2} a_{1}^{2}.
\end{eqnarray}
Then, Eqs.~\eqref{nu2eq} and~\eqref{h2eq} are cast into 
\begin{widetext}
\begin{eqnarray}
&&\label{y2eq} y_{2}'+\nu' h_{2} = \frac{\nu'}{2}e^{-\psi}a_{1}'^{2}+\frac{1}{3} \left[ \frac{e^{-\psi}}{r} \left( \nu'+\psi'+\frac{2}{r} \right) - \frac{2}{r^2}    \right] a_{1}a_{1}' + \frac{4\pi}{3} j_{1} \left(a_{1}'+\frac{2}{r}a_{1}\right),\\
\label{h2eqnew} 
&&h_{2}' + \frac{4e^{\psi}}{\nu' r^2} y_{2} + \left[ \nu' - \frac{8\pi e^{\psi}}{\nu'} (p_{0}+\rho_{0}) + \frac{2 }{r^{2} \nu'} (e^{\psi}-1)  \right]h_{2}=\frac{\nu'}{3}e^{-\psi}a_{1}'^{2} + \frac{4}{3r^2}a_{1}a_{1}'+\frac{16\pi}{3\nu' r^2} e^{\psi} j_{1}a_{1},\nonumber \\ 
\end{eqnarray}
\end{widetext}
where we used Eq.~\eqref{a1eqq} to eliminate $a_1''$. 

We now turn to the exterior problem and the boundary conditions. In the exterior region, Eqs.~\eqref{y2eq} and~\eqref{h2eqnew}
 can be solved analytically, yielding 
\begin{eqnarray}
\label{h2ext}
h_{2}^\mathrm{ext}(z) &=& K Q_{2}^{2}(z) +\hat{h}_{2}(z), \\ 
\label{y2ext} y_{2}^\mathrm{ext}(z) &=& -\frac{2K}{\sqrt{z^2 -1}} Q_{2}^{1}(z)+ \hat{y}_{2}(z) -\frac{e^{-\psi}}{6} a_{1}'^{2} \nonumber \\
&&-\frac{2e^{-\psi}}{3r}a_{1}a_{1}' -\frac{2}{3r^2} a_{1}^{2}.
\end{eqnarray}
Here, $K$ is an integration constant; $ Q_{b}^{a} $ denotes the second kind of associated Legendre functions
\begin{eqnarray}
Q_{2}^{2}(z) &\equiv& \frac{z(5-3z^2)}{z^{2}-1} + \frac{3}{2}(z^2 -1) \ln\bigg| \frac{z+1}{z-1} \bigg|, \\ Q_{2}^{1}(z) &\equiv& \frac{2-3z^2}{\sqrt{z^2 -1}} + \frac{3}{2} z(\sqrt{z^2 -1}) \ln \bigg| \frac{z+1}{z-1} \bigg|,
\end{eqnarray}
and
\begin{widetext}
\begin{eqnarray}
\hat{y}_{2}(z) &\equiv& \frac{3\mu^2}{8M^{4}}\frac{7z^2 -4}{z^2 -1} + \frac{3\mu^2}{16M^4} \frac{z(11z^2 -7)}{z^{2}-1} \ln\bigg|\frac{z-1}{z+1} \bigg| +\frac{3\mu^2}{16M^{4}} (2z^{2}+1) \ln\bigg|\frac{z-1}{z+1}\bigg|^2, \\
\hat{h}_{2}(z) &\equiv&-\frac{3\mu^2}{16M^4} \left[3z-\frac{2z(2z+1)}{z^2 -1}\right] -\frac{3\mu^2}{32M^4} \left[3z^2-8z-3-\frac{8}{z^2 -1}\right] \ln\bigg|\frac{z-1}{z+1} \bigg|\nonumber \\ && +\frac{3\mu^2}{16M^4} (z^2 -1) \ln\bigg|\frac{z-1}{z+1}\bigg|^2,
\end{eqnarray}
\end{widetext}
with $ z\equiv r/M -1. $ The interior and exterior solutions are matched at the stellar surface with the following boundary conditions:
\begin{eqnarray}
\label{eq:h2y2_BC}
[h_{2}]=0,\quad [y_{2}]=0.
\end{eqnarray}

\section{Numerical procedures}
\label{app:numerical}

We provide details on the numerical procedures for solving the interior equations derived in the main text and Appendix~\ref{app:pert_eqs}. 

\subsection{$\mathcal{O}(\epsilon^0)$: Background Spherically-symmetric Configurations}

Having these EoSs in hand, we solve the  interior equations at $\mathcal{O}(\epsilon^0)$. The mass equation in Eq.~\eqref{masseq}
 and the TOV equation in Eq.~\eqref{tov} are solved under the following boundary conditions at $r=r_0 \ll R$:
 \begin{eqnarray}
     m(r_0) = \frac{4\pi}{3}\rho_c r_0^3+ \mathcal{O}(r_0^5), \quad p(r_0) = p_c + \mathcal{O}(r_0^2),
 \end{eqnarray}
 where $\rho_c$ and $p_c$ are the central energy density and pressure respectively. The equation for $\nu$ in Eq.~\eqref{eq:nu}, on the other hand, is solved with the boundary condition at the surface in Eq.~\eqref{conditions}.

\subsection{$\mathcal{O}(\epsilon^1)$: Poloidal Magnetic Field Configuration}

We next explain the numerical procedure for solving the Maxwell equation in Eq.~\eqref{a1eqq} at $\mathcal{O}(\epsilon)$ with the current given in Eq.~\eqref{current}. Since this is a linear inhomogeneous equation, we can use the same technique for solving equations at second order in spin within the Hartle-Thorne formalism~\cite{Hartle:1967he,Hartle:1968si,Yagi:2013awa}. Namely, we split the full solution into the homogeneous $(a_{1h})$ and particular ($a_{1p}$) pieces
as
\begin{equation}
\label{a1lincom}
a_{1}(r) = a_{1p}(r) + c_{1} a_{1h}(r),
\end{equation}
with constant $c_1$. The boundary condition at $r=r_0$ is given by 
\begin{eqnarray}
    a_{1}(r_0)= a_{c} r_0^2 + \mathcal{O}(r_0^3), \quad  a_{1}'(r_0)= 2a_{c} r_0 + \mathcal{O}(r_0^2),  \nonumber \\
\end{eqnarray}
with constant $a_c$. Choosing a trial value for $a_c$, we first solve Eq.~\eqref{a1eqq} with $j_1=0$ to find a homogeneous solution $a_{1h}$. We next solve the full equation with the source for an arbitrary $c_0$ to find a particular solution $a_{1p}$. Having these in hand, we match the full solution for $a_1$ in Eq.~\eqref{a1lincom} with the exterior solution in Eq.~\eqref{a1ext} at the surface with the boundary condition in Eq.~\eqref{eq:a1_BC}, thereby obtaining $c_1$ and $\mu$ for a given $c_0$.

\subsection{$\mathcal{O}(\epsilon^2)$: Deformed Spacetime Configuration due to Magnetic Fields}

\label{secondorder}

We finally provide the numerical procedure for solving the interior equations in Eqs.~\eqref{y2eq} and~\eqref{h2eqnew} at $\mathcal{O}(\epsilon^2)$. 
Since these equations form a set of linear, inhomogeneous equations, we can adopt the numerical technique similar to that used at $\mathcal{O}(\epsilon)$. Namely, we once again split the solution into particular and homogeneous parts as
\begin{eqnarray} 
\label{h2y2part}
h_{2} = c_{2} h_{2}^{h} + h_{2}^{p},\quad
y_{2} = c_{2} y_{2}^{h} + y_{2}^{p},
\end{eqnarray}
for constant $c_2$. The particular solutions $h_2^p$ and $y_2^p$ are obtained by solving Eqs.~\eqref{y2eq} and~\eqref{h2eqnew} with the boundary condition
\begin{eqnarray}
\label{eq:h2y2_r0}
    h_2(r_0) &=& A r_0^2 + \mathcal{O}(r_0^3), \nonumber \\
    y_2(r_0) &=& \left[\left(-2\pi A+ \frac{16}{3}\pi a_{c}^{2}\right) \left(p_{c}+\frac{\rho_{c}}{3}\right) \right. \nonumber \\
   && \left. - \frac{4\pi}{3} a_{c}c_{0} (\rho_{c}+p_{c})\right] r_0^4 +  \mathcal{O}(r_0^5),
\end{eqnarray}
at $r=r_0$. $A$ is an integration constant that can be chosen arbitrary for constructing a particular solution. On the other hand, the homogeneous solutions $h_2^h$ and $y_2^h$ are obtained by solving Eqs.~\eqref{y2eq} and~\eqref{h2eqnew}  with $a_1=0$ (to switch off the source terms) and with the boundary condition in Eq.~\eqref{eq:h2y2_r0} with $a_c=0$. One then matches the full interior solutions in Eq.~\eqref{h2y2part}
 with the exterior solutions in Eqs.~\eqref{h2ext}  and~\eqref{y2ext} with the boundary conditions in Eq.~\eqref{eq:h2y2_BC} at the surface to determine the constants $c_2$ and $K$. The latter, in particular, allows us to estimate the stellar quadrupole moment and ellipticity.


\bibliography{references.bib} 

\end{document}